# A Linear Time Algorithm for Constructing Orthogonal Floor Plans with Minimum Number of Bends

Pinki[1]  and Krishnendra Shekhawat[1]

BITS Pilani, Department of Mathematics, Pilani Campus, India

**Abstract.** Let $\mathcal{G} = (V, E)$ be a *planar triangulated graph* (PTG) having every face triangular. A *rectilinear dual* or an *orthogonal floor plan* (OFP) of $\mathcal{G}$ is obtained by partitioning a rectangle into $|V|$ rectilinear regions (modules) where two modules are adjacent if and only if there is an edge between the corresponding vertices in $\mathcal{G}$.
In this paper, a linear time algorithm is presented for constructing an OFP for a given $\mathcal{G}$ such that the obtained OFP has $B_{min}$ bends, where a *bend* is a concave corner in an OFP. Further, it has been proved that at least $B_{min}$ bends are required to construct an OFP for $\mathcal{G}$, where $\rho - 2 \leq B_{min} \leq \rho + 1$ and $\rho$ is the sum of the number of leaves of the containment tree of $\mathcal{G}$ and the number of $\mathcal{K}_4$ (4-vertex complete graph) in $\mathcal{G}$.

**Keywords 1** algorithm, floor plan, graph theory, planar triangulated graph, separating triangle.

## 1. Introduction

In literature, there exist a lot of work related to *planar graphs* and its geometrical representation. A planar graph can be represented by a geometrical figure where vertices are replaced by geometrical objects and edges represent adjacency relation between the corresponding objects. This paper is about the representation of a planar graph in the form of a *floor plan* where vertices are represented by rectilinear polygons in such a way that the polygons corresponding to the adjacent vertices share a common boundary. We are specifically interested in the unweighted rectilinear representation of planar graphs which is known as *orthogonal floor plan* (OFP). In the weighted OFP, also known as *cartogram*, weights are associated with the vertices and are equal to the area of the corresponding polygon.

The floor planning problem is about constructing a floor plan while satisfying all adjacency requirements. Since it is always not possible to satisfy all adjacency constraints while maintaining rectangularity of a floor plan, it is often desirable to limit the number of bends in a floor plan for aesthetic, practical and cognitive reasons. In this work, we are dealing with planar triangulated graphs (PTG) and for a given PTG, we aim to construct an OFP with the minimum number of bends.

## 2. Preliminaries

Here, we present a few terminologies that are related to graph theory and floor plans which are frequently used in this paper. For further details refer to [1,2]. A *graph* is a set of vertices and edges denoted by $\mathcal{G}(V, E)$, where $n$ and $m$ denote the number of vertices and edges respectively. A graph is said to be *planar* if it can be embedded in the plane without crossing of edges. A *plane graph* is a planar graph with an embedding that divides the plane into connected components called *faces/regions*. The unbounded region is called *external face*. Except for the external face, all other faces are *internal faces*. $\mathcal{G}$ is called a *planar triangulated graph* (PTG) if all of its faces are triangle.

A *floor plan* (FP) is a closed polygon which is partitioned by straight lines into component polygons called *modules*. Two modules in a floor plan are *adjacent* if they share a wall or a section of it, where a *wall* of a module refers to the edges forming its perimeter. In a *rectangular floor plan* (RFP) the plan's boundary and each module are rectangles. For example, Figure 1a illustrates a planar graph $\mathcal{G}$ and its corresponding RFP. An *orthogonal floor plan* (OFP) has a rectangular plan boundary with a rectilinear partition in which the walls of each module are parallel to the sides of the plan boundary. For example, Figure 1b illustrates a planar graph for which a RFP does not exist, i.e, an OFP is required. A *bend* is a concave corner in an OFP (for an illustration, refer to Figure 1b).

### 2.1 Terminologies and Notations

**Definition 1** [ST] A *separating triangle* (ST) is a cycle of length 3 containing more than one vertex in its interior and if it contains only one vertex inside its interior then it is a 4-vertex complete graph $\mathcal{K}_4$.

For example: In Figure 2a, $\triangle 127$ is a $\mathcal{K}_4$ whereas $\triangle 124$ is a ST.

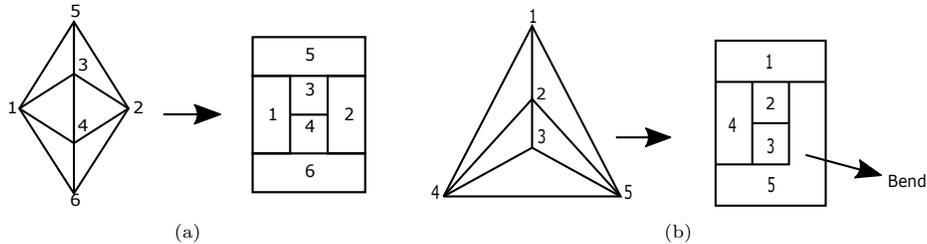

Fig. 1: Illustrating the concepts of the paper

**Definition 2** [Containment tree] A containment tree $T$ represents the hierarchy of the STs of a PTG. In $T$, each vertex represents a ST. The relation between two vertices is given by an edge. The hierarchical order of $T$ from root to leaves represents the containment of a vertex in its parent vertex. For a better understanding refer to Figure 3b.

**Definition 3** [Geometrical dual] The geometrical dual (with exterior face triangular) of a PTG $\mathcal{G}$ can be constructed as follows (refer to Figure 2):

1. Place a vertex of red color in each inner face of $\mathcal{G}$.
2. Place a vertex of blue color next to every edge of the outer face of $\mathcal{G}$ and call them $v_a$, $v_b$ and $v_c$.
3. If two inner faces are adjacent, join the corresponding red vertices with an edge in such a way that it only intersects the common edge between them.
4. If an inner face shares an edge with an outer face then join the corresponding red and blue vertices with an edge such that it only intersects the common edge between them.
5. Join $v_a$, $v_b$ and $v_c$ with an edge.

**Remark.**
The number of $\mathcal{K}_4$ in any PTG is equal to the number of $3-$cycles in its geometrical dual.

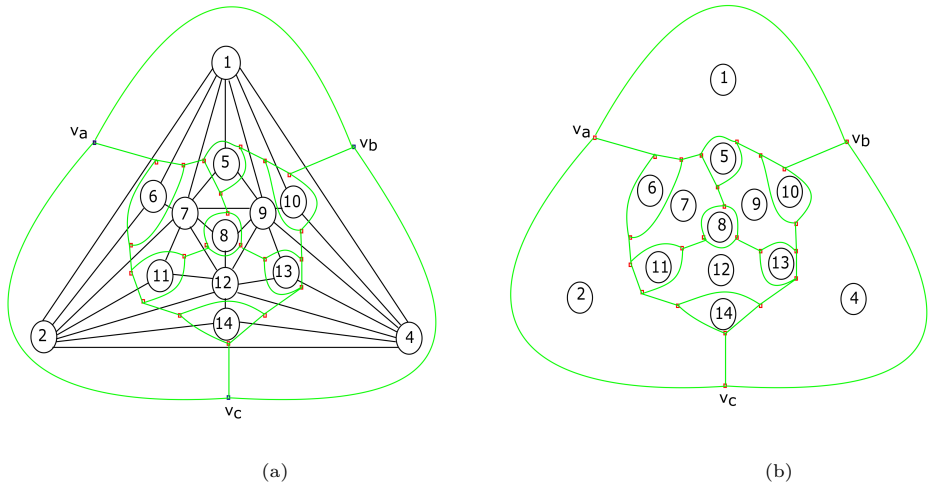

(a)          (b)

Fig. 2: (a) The construction of geometric dual (represented by green lines) of a subgraph $\mathcal{G}$ (having exterior face $\triangle 124$) of graph in Figure 3a (b) Geometric dual of $\mathcal{G}$

**Definition 4** [Grid Drawing] The grid drawing of a geometrical dual can be constructed as follows (refer to Figure 4a):

1. Consider the cycle say $C_1$ of shortest length with endpoints $v_a$ and not passing through $v_c$. Place all the vertices of $C_1$ on a cartesian plane horizontally from $v_a$ to $v_b$ with a distance of one unit between any two consecutive vertices and call it $H_1$ (horizontal level 1).

2. Similarly, consider the cycle say $C_2$ of shortest length with endpoints $v_a$ and not passing through $v_b$. Place all the vertices of $C_2$ vertically starting from a vertex which is adjacent to $v_a$ with a distance of one unit between any two consecutive vertices and call it $V_1$ ( vertical level 1).
3. Also, consider the cycle say $C_3$ of shortest length with endpoints $v_b$ and not passing through $v_a$. Place the vertices vertically starting from a vertex which is adjacent to $v_b$ (if not yet placed) at a distance of one unit.
4. Consider each vertex $v_i$ of $H_1$ and check if there is a vertex $v_j$ which is not yet covered and is adjacent to $v_i$ in its geometrical dual. If yes, place $v_j$ at a vertical distance of one unit to $v_i$. If it is not possible to add $v_j$ vertically, shift the vertex below to $v_i$ downwards by one unit and then place $v_j$.
5. Consider each vertex $v_i$ of $V_1$ and check if there is a vertex $v_j$ which is not yet covered and adjacent to $v_i$ in its geometrical dual. If yes, then place $v_j$ at a horizontal distance of one unit to $v_i$. If it is not possible to add $v_j$ horizontally, shift the vertex to the right of $v_i$ to the right by one unit and then place $v_j$.
6. Traverse through each horizontal and vertical level and place the uncovered vertices by following the steps 4 and 5.
7. Join all adjacent vertices (other than $v_a, v_b, v_c$) by horizontal, vertical, or diagonal line segments. Join $v_a, v_b, v_c$ by curved lines.

**Notations:**
FP: floor plan/s, RFP: rectangular FP, RFP($n$): RFP with $n$ modules, OFP: orthogonal FP, PTG: planar triangulated graph, PTPG: properly triangulated planar graph, ST: separating triangle, ST$^o$: ST as an exterior face, $k-$CRM: $k-$concave rectilinear module, $B_{min}$: minimum number of bends, CT: containment tree, $T_l$: leaves of containment tree, $M_j$: $j^{th}$ module of an OFP, $F_j$: $j^{th}$ inner face of geometrical dual of PTG.

## 3. Theoretical and Methodological Contributions

In our work, "planar graph" refers to a planar triangulated graph (PTG). In some of the literature mentioned here "planar graph" refers to a graph with all internal faces triangular and not containing any ST or $\mathcal{K}_4$ (also known as a properly triangulated planar graph (PTPG)). For any PTG, there does not exist a rectangular dual but there exists an OFP with the following properties:

1. Union of all polygons is a rectangle.
2. No four polygons can meet at a point.

Throughout this paper, RFP and OFP refer to the unweighted RFP and OFP.

### 3.1 History

In literature, there is a lot of work related to a class of planar graphs that allow rectangular duals or RFP. The problem of constructing a floor plan was first addressed for RFPs [3–8]. Then researchers proposed the construction of OFPs for the graphs for which RFP did not exist [9–15]. The brief literature related to RFP and OFP is as follows.

As an one of the initial work, in 1964, Levin [16] used adjacencies of a graph to address the floor plan layout problem. In 1985, Koźmiński and Kinnen [3] showed that a PTG admits a rectangular dual if and only if $\mathcal{G}$ has four vertices on its exterior and has no STs and $\mathcal{K}_4$. Simplifying the definition of PTG given in [3], in 1987 and 1988, Bhasker and Sahni [4, 5] proposed linear time algorithms for the existence and construction of RFP corresponding to PTPG. In 1988, Rinsma [6] provided conditions for the existence of RFP and an OFP for a given tree. In 1990, Lai and Leinwand [7] showed that the construction of a rectangular dual graph is equivalent to a matching problem in a bipartite graph derived from the given planar graph. In 1993, He [8] presented a new linear time algorithm for finding a RFP of PTPG, which is conceptually simpler than the previously known algorithms.

In 1993, Tsukiyama et al. [9] found that there does not exist an RFP for PTG because of the presence of ST and $\mathcal{K}_4$. He gave a heuristic algorithm for eliminating them and for constructing an RFP for the modified graphs in $O(n^2)$ time. In 1993, Sun and Sarrafzadeh [11] found that there exist PTG for which RFP does not exist and for those PTG, there may exist a floor plan having 0-CRM[1] and 1-CRM. In the same year, Yeap and Sarrafzadeh [10] proved that if only 0-CRM and 1-CRM are allowed, there are PTGs that do not admit a floor plan, i.e., 2-CRMs are sufficient and necessary for the feasibility of a floor plan. A linear-time algorithm for the construction of an OFP with 0-CRM, 1-CRM and 2-CRM was also proposed.

In 1999, He [12] presented a linear time algorithm that constructs an OFP for PTG using 0-CRM, 1-CRM and restricted 2-CRMs, i.e., *T*-modules and *Z*-modules (mirrored or rotated shapes are also considered). Using the concept of orderly spanning trees, in 2003, Liao et al. [13] gave a linear time algorithm for constructing an OFP for any *n*-vertex PTG which require fewer module types, i.e., the algorithm uses only *I*-modules, *L*-modules (could be flipped horizontally) and *T*-modules, but *Z*-modules are not required. In the same year, Kurowski [14] gave a simpler linear time algorithm for the construction of an OFP for PTG. He uses at most $1/2(n − 2)$ *T*-modules. In 2011, Jokar and Sangchooli [15] introduced the face area concept for constructing an OFP corresponding to a particular class of PTG.

A substantial amount of work done in literature involved bend number for an orthogonal drawing of planar graphs [17–19]. In an orthogonal drawing, vertices are represented by points and edges are represented by consecutive horizontal and vertical line segments. A bend is a point of intersection of edges. The orthogonal drawing did not ensure the rectangularity of the outer face of drawing.

In literature, a lot of work has also been done on cartograms [20–25]. Bends have also been discussed for them but in their context, bend number is defined as the least possible maximum complexity (i.e. the number of corners in a path) over all the paths for a given graph. It has also been proved that 8-sided rectilinear polygons are necessary and sufficient for the construction of weighted rectilinear duals for PTG.

---

[1] a module with *k* concave corners is said to be *k*-concave rectilinear module (*k*-CRM)

To the best of our knowledge, no one has talked about the minimum number of rectilinear modules that are required to address the adjacency constraints in a dual. In particular, none of the existing work talks about the number of bends in an OFP corresponding to a PTG. In this paper, we have proved that the minimum number of bends in an OFP lies between $\rho - 2$ and $\rho + 1$ where $\rho$ is equal to the sum of the number of leaves of the containment tree of $\mathcal{G}$ and the number of $\mathcal{K}_4$ in $\mathcal{G}$. Furthermore, we have proposed a linear time algorithm for the construction of an OFP for a given PTG with at most $\rho + 1$ bends.

## 4. Orthogonal Floor planning Algorithm for PTG

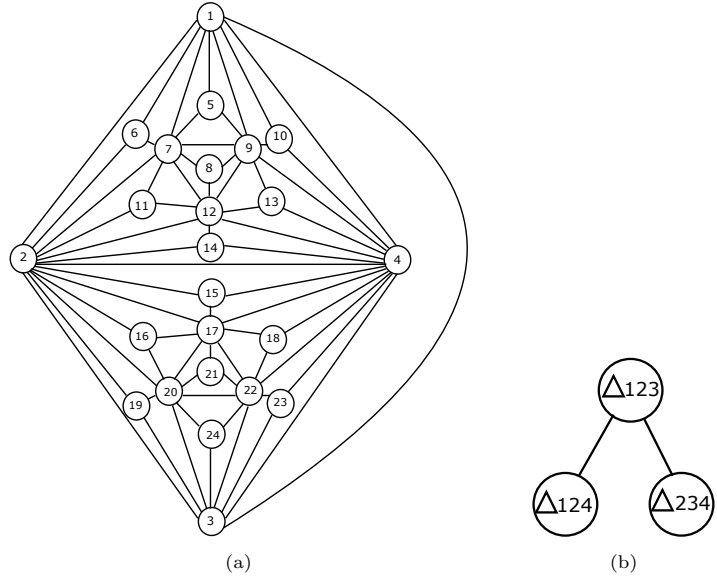

Fig. 3: (a) A PTG $\mathcal{G}$ with 3 ST and 14 $\mathcal{K}_4$ (b) Containment tree of $\mathcal{G}$

**Theorem 1.** $\rho - 2 \leq B_{min} \leq \rho + 1$ where $\rho = \# \, T_l + \# \, \mathcal{K}_4$.

*Case 1.*
(a) *If $\# \, ST \geq 2$ & no ST is contained in other ST except $ST^o$ then $B_{min} = \rho$.*
(b) *If $\# \, ST \geq 2$ & $\exists$ a $\mathcal{K}_4 \subseteq ST$ which do not shares an edge with any ST, then $B_{min} = \rho$.*

*Case 2.*

*If # $ST \geq 2$ & $\exists$ a $\mathcal{K}_4 \subseteq ST$ sharing an edge with the ST then $B_{min} = \rho - 1$.*

*Case 3.*
*If # $ST \geq 2$ & $\exists$ more than one ST containing $\mathcal{K}_4$ with a shared edge. If the shared edges have a common vertex then $B_{min} = \rho - 2$.*

*Case 4. If # $ST = 1$ and # $\mathcal{K}_4 \geq 1$ & $\exists$ a $\mathcal{K}_4 \subseteq ST$ sharing an edge or a vertex with the ST then $B_{min} = \rho$.*

*Case 5.*
*$B_{min} = \rho + 1$ if above cases do not hold.*

*Proof.*
**Case 5:** The existence of bends depends on triangles that are not a face, i.e., ST or $\mathcal{K}_4$. This implies that the contributors to bends are:

1. STs
   (a) $ST^o$ (the root of a CT): Every such triangle requires a bend to draw it in a rectangular shape, i.e., it contributes one to the number of bends.
   (b) STs (the leaves of a CT): Every leaf of a CT requires a bend for its rectangular representation. Hence, they contribute $\#T_l$ to the bends.
   (c) Intermediate STs of a CT: In a CT, intermediate STs are those STs which have a shared edge with either $ST^o$ or with STs that are leaves of CT. Because of the shared edge, the bend required for an intermediate ST is already covered by the ST with which it is sharing an edge. Hence, intermediate STs do not contribute to the bends.
2. $\mathcal{K}_4$s: Every $\mathcal{K}_4$ requires a bend for its rectangular representation. Hence, they contribute $\#\mathcal{K}_4$ to the bends.

Clearly, the minimum number of bends required to construct an OFP will be at most $\rho + 1 = \#T_l + \#\mathcal{K}_4 + 1$. Hence, $B_{min} = \rho + 1$, which can be reduced further as follows:

**Case 1(a):** # $ST \geq 2$ implies that $ST^o$ contains one or more STs in its interior and they further do not contain any ST, i.e., $ST^o$ always has a shared edge with the contained STs. Hence, a bend corresponding to any ST is also a bend for $ST^o$, i.e., a bend corresponding to $ST^o$ is not required and $B_{min} = \rho$.

**(b)** In this case, the presence of $\mathcal{K}_4$ does not affect the bends because it has no shared edge with STs. Therefore, only one bend is reduced because of $ST^o$ as discussed in Case 1(a). Hence, $B_{min} = \rho$.

**Case 2:** If a ST and a $\mathcal{K}_4$ are sharing an edge, then only one bend is required for both of them. The bend for $ST^o$ is also not required, as explained in Case 1(a). Hence, $B_{min} = \rho - 1$.

**Case 3:** Since a ST shares an edge with a $\mathcal{K}_4$, from Case 2, $B_{min} = \rho - 1$. Also, if more than one ST are sharing an edge with $\mathcal{K}_4$ and shared edges have a common vertex, then the bend at the common vertex gives bend to both triangles. Hence, $B_{min} = \rho - 2$.

**Case 4:** Since ST and $\mathcal{K}_4$ have a common edge or a common vertex, only one bend is required for both of them. Hence, $B_{min} = \rho$, i.e., $\#\mathcal{K}_4$.

**Example.** In Figure 3a, ST $\triangle 123$ contains two STs $\triangle 234$ and $\triangle 124$ and they contain many $\mathcal{K}_4 s$. The $\mathcal{K}_4$ $\triangle 24(12)$ shares an edge (2,4) with both STs. Similarly, ST $\triangle 234$ shares an edge (3,4) with $\mathcal{K}_4$ $\triangle 34(22)$. This is an example of Case 3 of Theorem 1. An OFP corresponding to this graph is shown in Figure 6b. We can easily see that the bend on module corresponding to a common vertex, i.e., $M_4$ is the required bend for both triangles $\triangle 24(12)$ and ST $\triangle 124$. Bend on $M_3$ is the bend for $\mathcal{K}_4$ $\triangle 34(22)$ and ST $\triangle 234$. One bend is reduced because of containment of $\triangle 234$ and $\triangle 124$ in ST$^o$ $\triangle 123$. The number of bends required for OFP in Figure 6b is 14.

**Corollary 1.** *If $\# ST = 1$ and $\# \mathcal{K}_4 = 0$ then $B_{min} = 1$.*

**Corollary 2.** *If $\# ST = 0$ and $\# \mathcal{K}_4 = 1$ then $B_{min} = 1$.*

**Remark.** The bend in $F$ appears on either 3-degree vertex or any vertex of separating triangle (ST).

---

**Algorithm 1** [OFP(G,T): Orthogonal Floor Planning Algorithm]

**Input**: A PTG $\mathcal{G}$ and its corresponding containment tree $T$
**Output**: An OFP with at most $\rho + 1$ bends

---

**Procedure:**
1: **For** $i = 1 : n$ where $n$ is the number of leaves of $T$ and every leaf corresponds a ST.
   Example: In Figure 3b, $T$ has two leaves $\triangle 124$ and $\triangle 234$.
2: **do**
3: Call OFP($\triangle ST$) corresponding to $i$.
4: **For** $i > 1$
5: **do**
6: Merge all $n$ obtained OFPs. Merging of two OFPs is done to the side which is common to both.
   Example: Figure 5b illustrates the merging of two OFPs that are given in Figure 5a.
7: Extend the module corresponding to any vertex of ST$^o$ which is not covered by the leaves of the containment tree $T$. Extend it with one unit length and breadth equals to the breadth of the obtained OFP. Extension is necessary to cover all adjacencies present in $\mathcal{G}$.
   Example: In Figure 5c, $M_1$ is extended to make it adjacent with $M_3$.

8: Bends can be further reduced by merging a rectangular part of outer module of obtained OFP with the outer module of any of its constituting OFPs. This can be accomplished only when a new adjacency is not created.
   Example: In Figure 6a, a part of $M_1$ is merged with $M_3$ and $M_{23}$ is extended horizontally in Figure 6b, to reduce a bend in the obtained OFP.
9: The desired OFP with $B_{min}$ bends is obtained.
   Example: In Figure 6b, an OFP with 14 bends, corresponding to $\mathcal{G}$ in Figure 3b, is illustrated.
10: **else** exit
11: **end**

    **Procedure:** OFP($\triangle ST$ )
1: Draw the geometrical dual to the given ST and call it $S$ (always 3-connected).
   Example: Figure 2b illustrates the geometrical dual of a sub-graph $\mathcal{G}$ (having exterior face $\triangle 124$) corresponding to $i = 1$.
2: Construct the grid drawing of $S$ and call it $S^*$ (see Definition 4).
   Example: Figure 4a illustrates $S^*$ for $S$ in Figure 2b.
3: Sort the inner faces of $S^*$ in the increasing order of their lengths and denote them by $F_j$ . List is not unique because faces of equal lengths are considered equally.
   Example: The sorted list of inner faces of $S^*$ in Figure 4a is $F_6, F_5, F_{11}, F_8, F_{10}, F_{13}, F_{14}, F_7, F_9, F_{12}$.
4: **For** $k = 1 : m$ where $m$ is the number of inner faces in $S^*$.
5: **do**
6: Identify an edge $e_p$ of $F_j$ which is neither horizontal nor vertical.
7: Break $e_p$ as a consecutive sequence of horizontal and vertical segment. It turns $F_j$ into a rectangular shape.
   Example: $F_6$ in Figure 4b is rectangular.
8: Replace the edges of the outer face which are not horizontal or vertical by a consecutive sequence of horizontal and vertical segments.
   Example: Refer to upper OFP in Figure 5a where edges on the exterior face in Figure 4b are expanded.
9: The desired OFP is obtained.
   Example: In Figure 5a, an upper OFP is obtained with 7 bends.

**Theorem 2.** *OFP($\mathcal{G}$,T) is well-defined and can be implemented to run in O(n) time.*

**Remark.** The area of any OFP is $W \times H$ (where $W$ and $H$ are the maximum number of modules connected horizontally and vertically) and the perimeter of $F$ is $2(W + H)$. For example, the OFP in Figure 6b has area 112 sq.units and perimeter 44 units. In Algorithm 1, an extra space is created in a module while merging two floor plans which will lead to an increase in the area of the OFP. The upper bound found for the area is not tight as it can be further reduced.

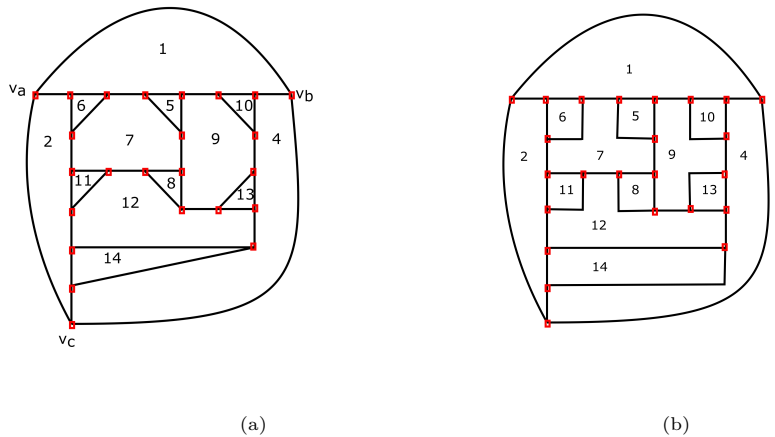

(a) (b)

Fig. 4: (a) Grid drawing $S^*$ obtained from geometrical dual given in Figure 2b (b) $S^*$ after making inner triangular faces rectangular

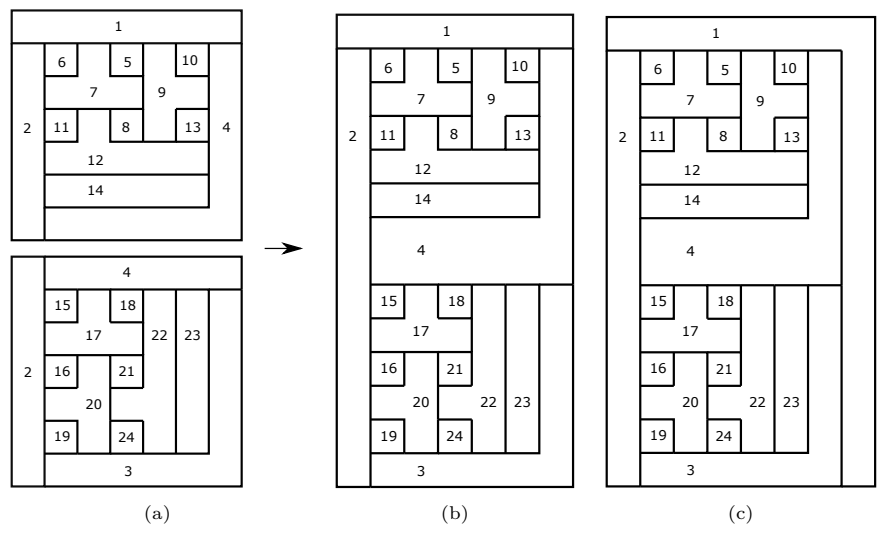

(a) (b) (c)

Fig. 5: (a) OFP for $\triangle 124$ and $\triangle 234$ (b) Merging of two OFPs of Figure 5a (c) Extension of $M_1$ to satisfy adjacency of $ST^o$ of $T$

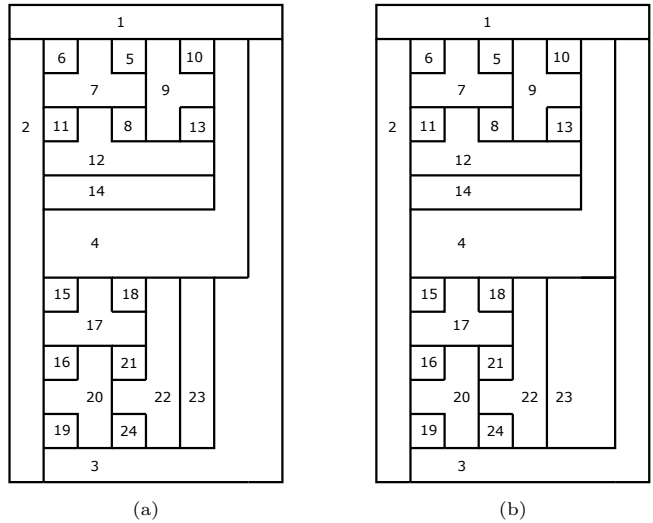

Fig. 6: (a) Merging of $M_3$ to $M_1$ (b) Reduction of a bend by extending $M_{23}$ and the desired OFP corresponding to $\mathcal{G}$ in 3a

## 5. Conclusion

Previously no researcher mentioned the number of bends of a floor plan in their work. Compared to the OFP obtained in ( [13,14,24]), our algorithm proved that the number of bends could further be reduced to a minimum number (refer to Figures 7 and 8). A linear time algorithm has been proposed for the construction of an OFP for a given PTG with at most $\rho+1$ bends. The algorithm is based upon a newly developed technique of containment tree which makes it simpler and efficient. Also, it is quite evident that our approach is practical and feasible. In future, we aim to construct OFP with a minimum area and rectilinear cartograms with a minimum number of bends.

## References

1. Jonathan L Gross and Jay Yellen. *Graph theory and its applications*. CRC press, 2005.
2. Ingrid Rinsma. Existence theorems for floorplans. *Bulletin of the Australian Mathematical Society*, 37(3):473–475, 1988.
3. Krzysztof Koźmiński and Edwin Kinnen. Rectangular duals of planar graphs. *Networks*, 15(2):145–157, 1985.
4. Jayaram Bhasker and Sartaj Sahni. A linear time algorithm to check for the existence of a rectangular dual of a planar triangulated graph. *Networks*, 17(3):307–317, 1987.
5. Jayaram Bhasker and Sartaj Sahni. A linear algorithm to find a rectangular dual of a planar triangulated graph. *Algorithmica*, 3(2):247–278, 1988.
6. I Rinsma. Rectangular and orthogonal floorplans with required room areas and tree adjacency. *Environment and Planning B: Planning and Design*, 15(1):111–118, 1988.

7. Yen-Tai Lai and Sany M Leinwand. A theory of rectangular dual graphs. *Algorithmica*, 5(1-4):467–483, 1990.
8. Xin He. On finding the rectangular duals of planar triangular graphs. *SIAM Journal on Computing*, 22(6):1218–1226, 1993.
9. Shuji Tsukiyama, Keiichi Koike, and Isao Shirakawa. An algorithm to eliminate all complex triangles in a maximal planar graph for use in vlsi floorplan. In *Algorithmic Aspects Of VLSI Layout*, pages 321–335. World Scientific, 1993.
10. Kok-Hoo Yeap and Majid Sarrafzadeh. Floor-planning by graph dualization: 2-concave rectilinear modules. *SIAM Journal on Computing*, 22(3):500–526, 1993.
11. Yachyang Sun and Majid Sarrafzadeh. Floorplanning by graph dualization: L-shaped modules. *Algorithmica*, 10(6):429–456, 1993.
12. Xin He. On floor-plan of plane graphs. *SIAM Journal on Computing*, 28(6):2150–2167, 1999.
13. Chien-Chih Liao, Hsueh-I Lu, and Hsu-Chun Yen. Compact floor-planning via orderly spanning trees. *Journal of Algorithms*, 48(2):441–451, 2003.
14. Maciej Kurowski. Simple and efficient floor-planning. *Information processing letters*, 86(3):113–119, 2003.
15. Mohammad Reza Akbari Jokar and Ali Shoja Sangchooli. Constructing a block layout by face area. *The International Journal of Advanced Manufacturing Technology*, 54(5-8):801–809, 2011.
16. Peter Hirsch Levin. Use of graphs to decide the optimum layout of buildings. *The Architects' Journal*, 7:809–815, 1964.
17. Takao Nishizeki and Md Saidur Rahman. *Planar graph drawing*, volume 12. World Scientific Publishing Company, 2004.
18. Md Saidur Rahman, Kazuyuki Miura, and Takao Nishizeki. Octagonal drawings of plane graphs with prescribed face areas. *Computational Geometry*, 42(3):214–230, 2009.
19. Siddharth Bhatia, Kunal Lad, and Rajiv Kumar. Bend-optimal orthogonal drawings of triconnected plane graphs. *AKCE International Journal of Graphs and Combinatorics*, 15(2):168–173, 2018.
20. Marc Van Kreveld and Bettina Speckmann. On rectangular cartograms. *Computational Geometry*, 37(3):175–187, 2007.
21. Mark De Berg, Elena Mumford, and Bettina Speckmann. On rectilinear duals for vertex-weighted plane graphs. *Discrete Mathematics*, 309(7):1794–1812, 2009.
22. Torsten Ueckerdt. Geometric representations of graphs with low polygonal complexity. *thesis*, 2012.
23. David Eppstein, Elena Mumford, Bettina Speckmann, and Kevin Verbeek. Area-universal and constrained rectangular layouts. *SIAM Journal on Computing*, 41(3):537–564, 2012.
24. Md Jawaherul Alam, Therese Biedl, Stefan Felsner, Michael Kaufmann, Stephen G Kobourov, and Torsten Ueckerdt. Computing cartograms with optimal complexity. *Discrete & Computational Geometry*, 50(3):784–810, 2013.
25. Yi-Jun Chang and Hsu-Chun Yen. Rectilinear duals using monotone staircase polygons. In *International Conference on Combinatorial Optimization and Applications*, pages 86–100. Springer, 2014.

## 6. Appendix

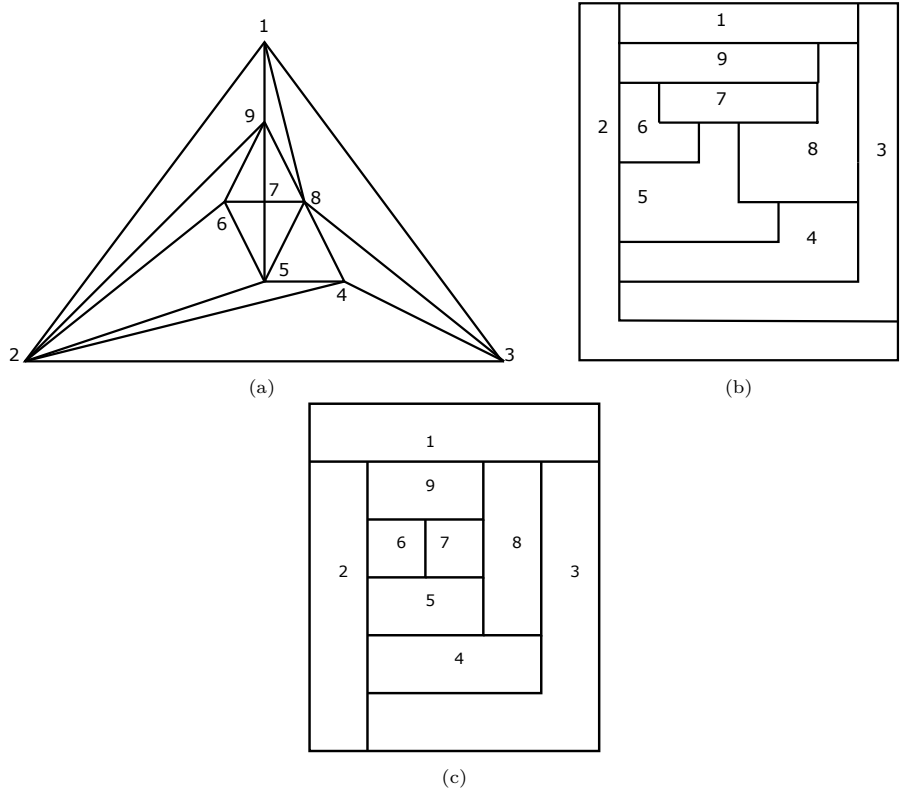

Fig. 7: (a) A PTG $\mathcal{G}$ from Alam [24] (b) OFP corresponding to $\mathcal{G}$ given in [24] has 7 bends with area = 72 sq.unit and perimeter = 34 unit (c) In comparison to (b) OFP obtained by Algorithm 1 has one bend with area = 30 sq. unit and perimeter = 22 unit

**Theorem 3.** *OFP($\mathcal{G}$,T) is well-defined and can be implemented to run in $O(n)$ time.*

*Proof.* The correctness of Algorithm 1 is given as follows:

1. **Precondition:** A containment tree $T$ of $\mathcal{G}$ whose vertices represent STs and edges in $T$ represents the adjacency in the corresponding OFP.
2. **Postcondition:**
   An OFP with at most $\rho + 1$ bends, where $\rho = \#\, T_l + \#\, \mathcal{K}_4$. The area and perimeter of the obtained OFP are $W \times H$ and $2(W + H)$ respectively.

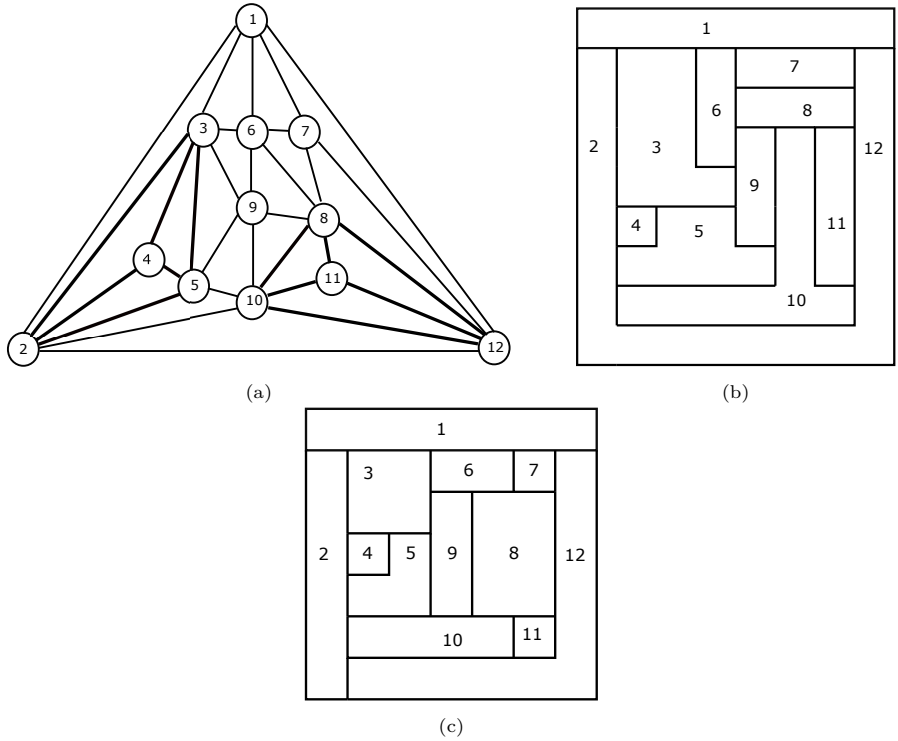

Fig. 8: (a) A PTG $\mathcal{G}$ from Liao [13] (b) OFP corresponding to $\mathcal{G}$ given in [13] has 6 bends with area = 72 sq.unit and perimeter = 34 unit (c) In comparison to (b) OFP obtained by Algorithm 1 has one bend with area = 49 sq. unit and perimeter = 28 unit

**Explanation:** In Algorithm 1, we are constructing a geometrical dual corresponding to every leaf of the containment tree $T$. In the geometrical dual of a ST, it is represented by an outer cycle and contained $\mathcal{K}_4s$ are represented by 3-cycles. Every 3-cycle can be drawn in the form of a rectangular face with one bend. Similarly, for maintaining the adjacencies in the corresponding OFP, the outer face corresponding to $ST^o$ is drawn with one bend. Hence, the number of bends needed to draw an OFP corresponding to a ST is the number of bends for $\mathcal{K}_4s$ (3-cycle) along with a bend for $ST^o$ which equals the number of $\mathcal{K}_4s$ +1.

If there are $n$ leaves where every leaf is a ST, then the number of bends is the sum of the number of STs and the number of $\mathcal{K}_4s$. In Algorithm 1, we are constructing an OFP corresponding to every leaf of $T$ except the root. By merging all corresponding OFPs, we obtain an OFP for $G$ in which adjacency corresponding to $ST^o$ is not covered, which requires one more bend. Hence, the number of bends in an obtained OFP is the sum of the number of STs, the number of $\mathcal{K}_4s$ and one bend for $ST^o$.

Every adjacency between modules in an OFP is considered as one unit. The area of an OFP is the product of the maximum number of horizontal and vertical connections which equals $W \times H$. Similarly, the perimeter is $2(W + H)$.

The step-wise explanation of the implementation of Algorithm 1 in linear time is as follows:

**Step 1-3**: The construction of a geometrical dual and a grid drawing can be done easily in a constant time. To obtain the required OFP, we need to draw every inner face of length 3 in the form of a rectangle which again can be done in a constant time. Since the for-loop of this step is from $k = 1$ to $m$, it runs in a linear time.

**Step 6**: This step can be implemented in O($n$) time because the merging of two OFPs takes O(1) time. Hence, the merging of $n$ OFPs takes linear time.

**Step 7-11**: To satisfy all adjacencies, we need to take care of the adjacency of ST$^o$ which is not considered yet. That can be done by extending either length or breadth of the outer module by one unit. Extending and reducing the rectangular space takes O(1) time, hence, this step runs in a constant time.